# Sculpting oscillators with light within a nonlinear quantum fluid

G. Tosi, G. Christmann, N.G. Berloff, P. Tsotsis, T. Gao, Z. Hatzopoulos, P.G. Savvidis, J.J. Baumberg

**Seeing macroscopic quantum states directly remains an elusive goal. Particles with boson symmetry can condense into such quantum fluids producing rich physical phenomena as well as proven potential for interferometric devices[1,2,3,4,5,6,7,8,9,10]. However direct imaging of such quantum states is only fleetingly possible in high-vacuum ultracold atomic condensates, and not in superconductors. Recent condensation of solid state polariton quasiparticles, built from mixing semiconductor excitons with microcavity photons, offers monolithic devices capable of supporting room temperature quantum states[11,12,13,14] that exhibit superfluid behaviour[15,16]. Here we use microcavities on a semiconductor chip supporting two-dimensional polariton condensates to directly visualise the formation of a spontaneously oscillating quantum fluid. This system is created on the fly by injecting polaritons at two or more spatially-separated pump spots. Although oscillating at tuneable THz-scale frequencies, a simple optical microscope can be used to directly image their stable archetypal quantum oscillator wavefunctions in real space. The self-repulsion of polaritons provides a solid state quasiparticle that is so nonlinear as to modify its own potential. Interference in time and space reveals the condensate wavepackets arise from non-equilibrium solitons. Control of such polariton condensate wavepackets demonstrates great potential for integrated semiconductor-based condensate devices.**

Non-resonant optical pumping of a semiconductor microcavity in the strong coupling regime, continuously injects incoherent carriers which rapidly cool and scatter into the mixed light-matter states known as polaritons. Above a certain pump threshold (10mW at $T$=10K) these polaritons Bose condense [11,12,13]. Because of the extreme nonlinearities caused by strong repulsion between polaritons, they are shifted to higher energies (blue-shifted) wherever the density is high, particularly at the pumped spot[17,18]. The polaritons thus feel an outward force and form an expanding polariton condensate[19]. In this paper we explore the novel effects that occur when two neighbouring polariton condensates interact. Instead of typical Josephson-junction coherent coupling phenomena[20], new effects arise because of the quasiparticle interactions.

The decreasing density (and hence blue-shifts) away from two pumped spots induces a two-peaked potential profile (Fig.1a). Surprisingly, on the line between the pump spots this potential appears parabolic forming a potential trap like that of a simple harmonic oscillator, the quantum equivalent of a pendulum. The polaritons experiencing this potential redistribute in energy and space to occupy the simple harmonic oscillator (*SHO*) states (Fig.1b). Because the polaritons can slowly leak through their confining mirrors into escaping photons, spatial images recorded on a camera (filtered at each emission energy) directly reveal the characteristic quantum wavefunctions (Fig.1c), tens of microns across. Such extended coherent quantum states in a semiconductor are unprecedented to image in real time directly. The energy spacing between levels (Fig.1d) is almost identical, $E_n = \hbar v(n_{SHO} + \frac{1}{2})$, and the increasing number of spatial nodes with $n_{SHO}$ are clearly resolved . Along the line between pump spots, the wavefunctions fit very well the expected Hermite-Gaussian $\psi_{SHO}(x)$ states (Fig.1e). From such fits, the polariton potential is reconstructed (see Supp.Info.) as superimposed in Fig.1(b).

**Figure 1: Spatially-mapped polariton condensate wavefunctions**

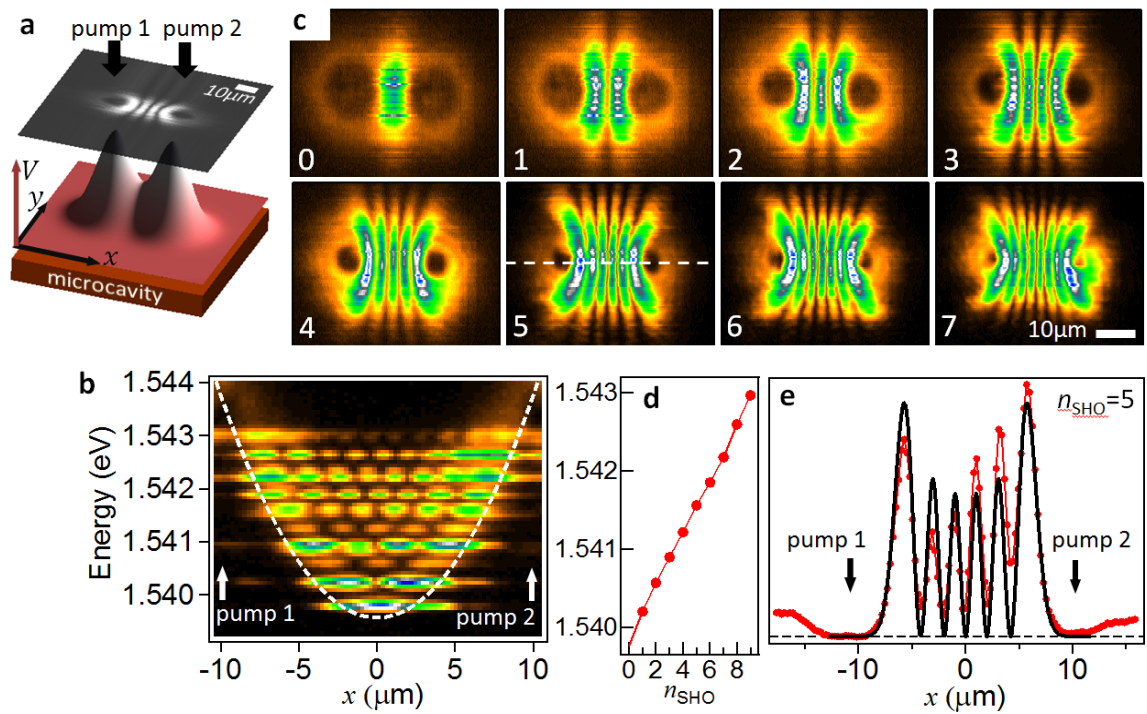

**a**, Expt scheme with two 1μm-diameter pump spots of separation 20μm focussed onto the planar microcavity. The effective potential $V$ (red) produces multiple condensates (grey image shows $n_{SHO}$=3 mode). **b**, Real space spectra along line between pump spots. **c**, Tomographic images of polariton emission (repulsive potential seen as dark circles around pump spots). Labelled according to $n_{SHO}$ assigned from **d**, extracted mode energies vs quantum number. **e**, Hermite-Gaussian fit of $\psi_{SHO}^{n=5}(x)$ to image cross section, dashed in **c**.

By controlling the spacing between the pump spots the shape and orientation of this *SHO* potential, $V(\mathbf{r})$, can be directly modified in real time, thus changing the energy level spacing $\hbar\nu$ (Figs.2a,b). By contrast if only one pump is present, the condensate polaritons flow out unhindered without relaxation (thus remaining at the blue-shifted polariton energy at the pump spot, Fig.2c), while below threshold only incoherent emission is observed around each pump spot. Plotting the quantised energy levels for several pump separations, $L$, confirms their equal energy spacing, and the predicted inverse dependence on separation (Fig.2d). Although recent 1D versions of coherent polariton states observed in microcavities etched into wires[19] do not appear to possess energies linked to any spatial scales, we suggest they arise in a similar fashion to here. As we show below, in 2D other significant features arise, including periodic oscillations of the polariton wavepackets and control of the polariton potential (Supp. Info.). Our crucial advance is the ability to manipulate independent condensates in 2D on a chip, using externally imprinted potentials which are not statically predefined.

If the pump separation is kept constant but the pump powers are both increased (Supp. Info. Figs.S2a,b), then the spacing $\hbar\nu$ only increases slightly (Fig.S2c,d). However the increasingly-deep potential (from the pump-induced blue-shifts) increases the number of *SHO* states trapped inside. From the *SHO* model increasing the polariton density, $|\psi|^2$, gives a linear blue-shift, $V_{max} = g|\psi|^2$ (Fig.S2d, right axis) producing the dependence as observed,

$$\hbar\nu = \sqrt{\frac{2\hbar^2}{m^*} \cdot \frac{\partial^2 V}{\partial x^2}} = \sqrt{\frac{2\hbar^2}{m^*} \cdot \frac{V_{max}}{L^2}} \simeq \frac{\hbar}{L}\sqrt{\frac{2g|\psi|^2}{m^*}}. \qquad (1)$$

**Figure 2: Dependence of simple harmonic oscillator states on pump properties**

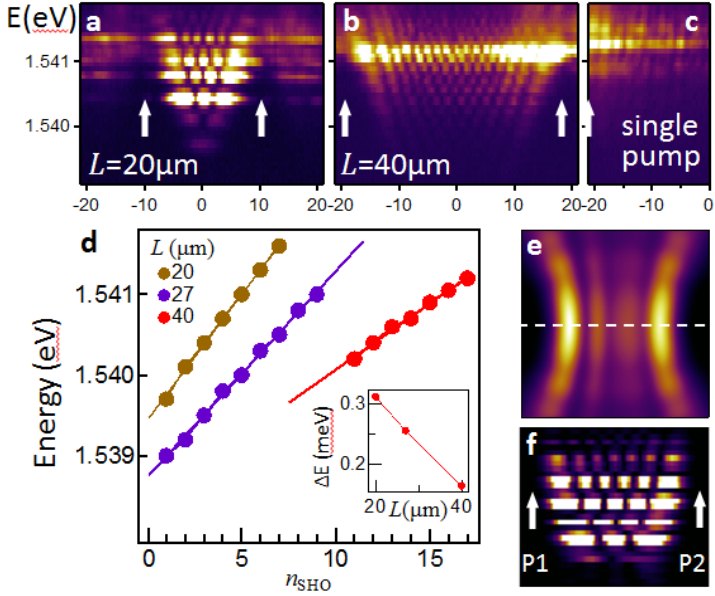

**a-c**, Spatially-resolved polariton energies on a line between pump spots (white arrows). In **a-b**, the pump separation controls the *SHO* energy spacing. In **c**, no relaxation is observed. **d**, Ladder of *SHO* energies for 3 different pump separations, with average energy spacings extracted in the inset. **e**, Time-averaged 2D simulation of cGL equation for 20µm pump separation and **f**, resulting time-averaged spectra along dotted line in **e**.

The *SHO* states are thus only resolved due to the strong polariton repulsion ($g$) and the ultra-light polariton mass, $m^* = 4.2 \times 10^{-5} m_e$, (measured independently, see Supp.Info). The polariton *SHO* states are populated differently under different conditions, with high pump powers and close pump separations favouring relaxation to the lower condensate *SHO* states. By using non-equal pump powers asymmetric potentials can also be created (see Supp.Info.).

Our theoretical explanation starts with the mean field equation[21,22,23] for the lower polariton wavefunction, $\psi$, in the presence of reservoir population, $N$, of optically-injected hot excitons

$$i\hbar\frac{\partial\psi}{\partial t} = [E(i\nabla) + g|\psi(r,t)|^2 + V_{ext}(r)]\psi + i\frac{\hbar}{2}[P(r,t,N) - \Gamma_C]\psi \qquad (2)$$

which includes the polariton dispersion $E(k) = \hbar^2 k^2/2m^*$, the strength of the contact interaction potential $g$, an external potential $V_{ext}(r)$ that describes repulsive interactions with the reservoir, the rate of polariton losses $\Gamma_C$, and the incoherent pump $P$ (details are given in the supplementary information). The net polariton potential $V(r) = g|\psi(r,t)|^2 + V_{ext}$ is produced by the repulsive interactions between polaritons themselves as well as with the reservoir excitons ($N$) close to the pump spots. This extension of the Gross-Pitaevskii equation is a complex Ginzburg-Landau (cGL) equation, a universal equation of mathematical physics describing the behaviour of systems in the vicinity of an instability and symmetry breaking[24] and capable of spontaneous pattern formation. The non-equilibrium solution arises from the constant localised energy input at the pump spots together with polariton decay. Relaxation of polaritons by acoustic phonon emission is very slow[25], and swamped by polariton-polariton and polariton-exciton scattering over the timescales discussed below. For two polaritons initially in $SHO$ states $n_1, n_2$ their mutual scattering to states $n_1', n_2'$ is only energy-phase matched if $n_1' + n_2' = n_1 + n_2$ and $E_1' - E_1 = -\{E_2' - E_2\}$. Scattering is thus most rapid if the energy separations between states are equal. At low powers, the potential is not parabolic giving unequal energy spacings and thus scattering is slower. At higher polariton densities, the scattering rate increases so that polariton relaxation populates lower energy states. The resulting new polariton density profile modifies the polariton potential leading to more parabolic and equally spaced energies, thus speeding up scattering and feeding back positively. This contrasts to current theories for polariton condensates postulating a phonon scattering energy threshold[21] (that we see no evidence for in experiment). The self-organised nature of the highly nonlinear polaritons is to produce the most rapid non-equilibrium energy flow through the system, by forming an $SHO$ parabolic potential with an energy ladder that maximises polariton relaxation. To our knowledge this is the first solid state quasiparticle that is so nonlinear as to modify its own potential in this way, and links to current theories of nonequilibrium systems.[24]

We now demonstrate that as well as the individual $SHO$ states being coherent condensates, they are also mutually coherent with each other. We compare first with the nonlinear Maxwell's equations for optical pulses propagating inside a nonlinear medium with susceptibility $\chi$ so that $i\dot{E} = \{\chi^{(1)} + \chi^{(3)}|E|^2\}E$. While propagating optical pulses generate equally-spaced sidebands through four wave mixing[26], condensate polaritons analogously parametrically scatter to give new coherent polariton states. In optics, such nonlinear dispersions can produce solitons and lead to mode-locking of laser cavity modes to produce trains of coherent intense pulses.[26] Here polaritonic wavepackets appear spontaneously bouncing back and forth between the two pump spots, corresponding exactly to the spectral and spatial organisation observed.

The real space images (not energy filtered) are split in a Michelson interferometer and recombined at a CCD camera (Fig.3a-c), with the fringes aligned in the vertical direction. We map the fringe visibility and hence first order coherence $g^{(1)}(r, r, t)$ at the same spatial locations but separated in time (Fig.3a-c, lower). As the time delay between the interfering images increases, the fringe visibility rapidly decreases everywhere but revives strongly after $t_r = \pm 13$ps (Fig.3c,d). This revival implies not only that the individual $SHO$ states are phase coherent but that they are condensates with a stable phase relationship.

All observations above are accounted for by a single condensed coherent state, $\psi$, describing a polariton wavepacket in the quantum liquid oscillating back and forth with period $t_r$ producing the characteristic $SHO$ sidebands observed. The oscillation period scales as expected with pump spatial separation (Fig.S5). Indeed from equation (1), $t_r = \pi L \sqrt{m^*/2g|\psi|^2}$ (plotted as open circles), again consistent with the wavepacket oscillations. These 0.3-1.0THz $SHO$ sidebands, generated by a propagating condensate wavepacket moving through its own nonlinear potential, correspond to the frequency micro-combs recently observed in cavity optomechanics[27].

**Figure 3: Coherence revivals in real-space condensate wavepacket interferometry**

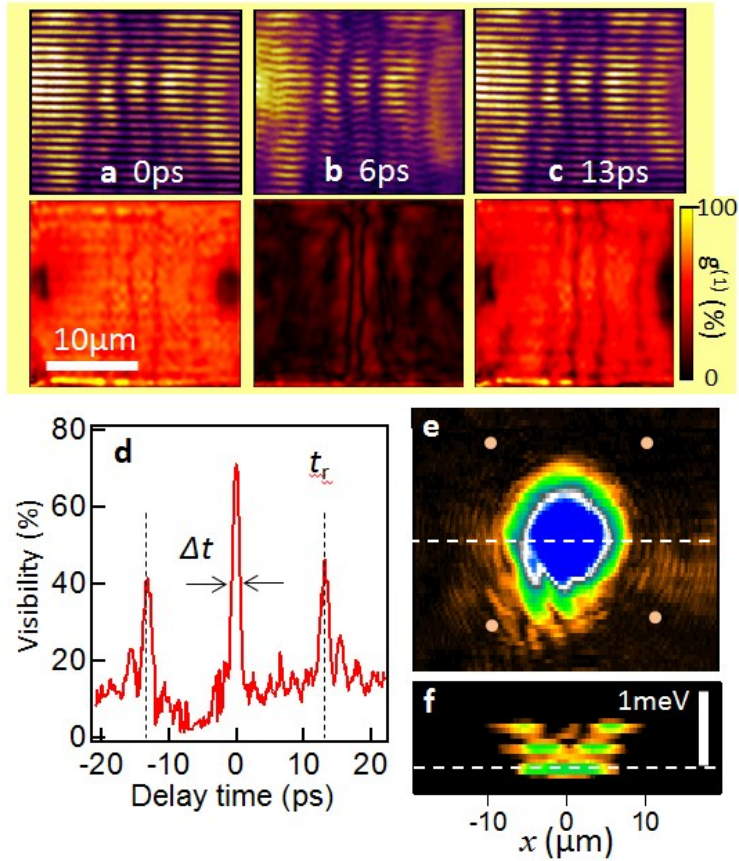

**a-c**, Real space interference pattern for time delays marked. Pump spots separated by 20μm, just off right and left side of image. Lower images show extracted first order coherence, $g^{(1)}(r,r,t)$. **d**, Fringe visibility averaged over each image *vs* Michelson time delay. **e**, Real space tomographic image of *n*=1 state for 4 pump spots (marked, with actual spot size), with **f**, energies across dashed line.

The revival fringe amplitude at increasing time delays decays with a 40ps lifetime, due to coherent wavepacket dispersion, decay, dephasing and diffusion. This lifetime is however in good agreement with the coherent lifetime of the condensate found by converting the decay length of the ballistically expanding condensate into time. Wavepacket dispersion arises from the variation in the energy level spacing, $\Delta(\hbar v)_{\text{rms}}$ ~50μeV, contributing 80ps to the coherence decay. The temporal width of the condensate wavepacket corresponds to the number of $SHO$ states observed, $\Delta t \simeq t_r/n_{SHO}$, (with $n_{SHO}$~10 well above threshold). In this picture, when the interacting condensates phase lock, the resulting spatially-modulated polaritons create polariton sidebands at the $SHO$ states. In turn this further sharpens up the wavepacket extent and enhances the nonlinear polariton scattering. Such condensate polariton solitary waves are expected in cGL equations, and resemble spatial solitons recently observed in atomic BECs[28,29]. Higher-energy sidebands escape the $SHO$ potential (unlike atomic BECs) leading to a modified solitary wave structure (note dark solitons are expected for the perfect 1D parabolic potential[30]). The behaviour observed is matched by our simulations of the cGL equation (Supp. Info.). Multiple wavepackets are observed bouncing back and forth (Fig.2e, Fig.S4 and movie in Supp.Info.) which indeed produce spectra corresponding to the $SHO$ sidebands (Fig.2f).

This oscillating quantum liquid model explains another otherwise peculiar feature. While polaritons are well confined in between the pump spots, they are unconfined and even ejected by the potential (Fig.1a) in the perpendicular direction. In spite of this, well defined $SHO$ wavefunctions (Fig.1) are seen, which would not be expected for non-interacting 2D quantum quasiparticles. This is explained by the oscillating wavepacket, which is repeatedly amplified close to each pump spot on each pass.

The flexibility of the pump-induced polariton condensates easily allow full 2D confinement using extra pump spots. For instance, the lowest energy state in the 4-square pump spot potential is fully confined in the $xy$ plane (Fig.3e,f), and selective polariton condensate beams can be extracted. Thus polariton condensate circuits can now be created on the fly merely by appropriate sculpting of the pumping geometry, which will lead to many future developments.

**Methods:**

To produce the effects seen here, which persist all across the microcavity samples, high quality growth is required. A 5λ/2 AlGaAs DBR microcavity is used for all experiments with four sets of three quantum wells placed at the antinodes of the cavity electric field[31]. The cavity quality factor is measured to exceed Q>8000, with transfer matrix simulations giving $Q=2\times10^4$ corresponding to a cavity photon lifetime T=9ps. Strong coupling is obtained with a characteristic Rabi splitting between upper and lower polariton energies of 12 meV. The microcavity wedge allows scanning across the sample to set the detuning between the cavity and photonic modes. All data presented here use a negative detuning of -5meV, although other negative detunings give similar results. Excitation is provided by a single-mode narrow-linewidth CW laser, focussed to 1µm diameter spots through a 0.7 numerical aperture lens, and tuned to the first spectral dip at energies above the high-reflectivity mirror stopband at 750nm. To prevent unwanted sample heating the pump laser is chopped at 100Hz with a on:off ratio of 1:30.

The sample is held at cryogenic temperatures below 10K, though similar effects are seen at higher temperatures. Images are recorded on an uncooled Si CCD camera in the magnified image plane, while spectra are recorded through a 0.55m monochromator with liquid nitrogen cooled CCD. Tomography uses a computer controlled mirror selecting the line illumination of the front aperture of the monochromator. The Michelson uses retroreflectors mounted on delay stages to control the relative temporal separation of the interference. The pump laser is spectrally filtered out of all images. The fringe visibility and hence first order coherence is mapped by extracting the 1st order diffraction from the Fourier transformed images produced in the image plane after the Michelson. These are normalised to the corresponding 0th order diffraction and then back-transformed into real space images.

Affiliations

**NanoPhotonics Centre, Cavendish Laboratory, Department of Physics, JJ Thompson Ave, University of Cambridge, Cambridge, CB3 0HE, UK**

G. Tosi*, G. Christmann, J.J. Baumberg

* also at **Departamento de Física de Materiales, Universidad Autonóma, E28049 Madrid, Spain**

**Department of Applied Mathematics and Theoretical Physics, University of Cambridge, Cambridge, CB3 0WA**

N.G. Berloff

**Department of Materials Science & Technology,**
**Department of Physics, University of Crete, PO Box 2208, 71003 Heraklion, Crete, Greece**
**Foundation for Research and Technology-Hellas, Institute of Electronic Structure & Laser, PO Box 1527, 71110 Heraklion, Crete, Greece**

P. Tsotsis, T. Gao, Z. Hatzopoulos, P.G. Savvidis,

**Correspondence and requests** for materials should be addressed to j.j.baumberg@phy.cam.ac.uk

**Author Contributions**: G.T. and G.C. performed the spectroscopy experiments, and together with J.J.B. analysed the data and wrote the manuscript. P.G.S. contributed to the preparation of the manuscript and together with P.T., T.G., Z.H. designed and grew the microcavity samples, providing characterisation spectroscopy to sustain high quality performance. N.G.B. devised, coded, and carried out the modelling simulations.

**Supplementary Information** is linked to the online version of the paper at www.nature.com/nature.

**Acknowledgements** to L. Vina for comments, and grants EPSRC EP/G060649/1, EU CLERMONT4 235114, EU INDEX 289968, Spanish MEC (MAT2008-01555) and Greek GSRT program Irakleitos II. G.T. acknowledges financial support from an FPI scholarship of the Spanish MICINN.

**Competing financial interests statement.**
The authors declare no competing financial interests.